\documentclass[a4paper]{article}
\usepackage{graphicx}      
\usepackage{amsmath,amssymb,amsfonts}
\usepackage{subfigure}
\usepackage[pagebackref=false]{hyperref}
\newcommand{\interior}[1]{\overset{\smash{\raisebox{-0.05ex}{$\scriptstyle\circ$}}}{#1}\rule{0pt}{2.ex}}
\def \asinh {\text{asinh}}

\begin{document}

\title{Essentially high-order compact schemes with application to stochastic volatility models on non-uniform grids}

\author{Bertram D{\"u}ring\thanks{Department of Mathematics,
    University of Sussex, Pevensey II, Brighton, BN1 9QH, United
    Kingdom, \texttt{bd80@sussex.ac.uk}} \and Christof Heuer\thanks{Chair of Applied Mathematics / Numerical Analysis,
  Bergische Universit{\"a}t Wuppertal, Gau\ss stra\ss e 20, 42097 Wuppertal,
Germany, \texttt{cheuer@uni-wuppertal.de}}}

\maketitle
\begin{abstract}
\noindent We present high-order compact schemes for a linear second-order
parabolic partial differential equation (PDE) with mixed second-order
derivative terms in two spatial dimensions. The schemes are applied to
option pricing PDE for a family of stochastic volatility models.
We use a non-uniform grid with more grid-points around the strike price. 
The schemes are fourth-order accurate in space and second-order accurate in time for vanishing correlation. 
In our numerical convergence study we achieve fourth-order accuracy also for non-zero correlation. 
A combination of Crank-Nicolson and BDF-4 discretisation is applied in
time. Numerical examples confirm that a standard, second-order finite
difference scheme is significantly outperformed.  
\end{abstract} 
\section{Introduction}
We consider the following parabolic partial differential equation for\\$u~=~u(x_1,x_2,t)$ in two spatial dimensions and time,
\begin{align}\label{Definition_of_general_grid_2D_EHOC_before_using_f}
du_{\tau} + a_1u_{x_1x_1} + a_2 u_{x_2x_2} + b_{12} u_{x_1x_2} +
  c_1u_{x_1} + c_2  u_{x_2}=0 \quad \text{in } \Omega \times ]0,T]=:Q_T,
\end{align}
subject to suitable boundary conditions and initial condition
$u(x_1,x_2,0) = u_0(x_1,x_2)$ with $T >0$ and   
$\Omega = \bigl[x_{\min}^{(1)} ,x_{\max}^{(1)}  \bigr] \times
\bigl[x_{\min}^{(2)} ,x_{\max}^{(2)}  \bigr] \subset\mathbb{R}^2$
with $x_{\min}^{(i)} <x_{\max}^{(i)}$ for $i=1,2$.  
The functions $a_i=a_i(x_1,x_2,\tau)<0$,
$b_{12}=b_{12}(x_1,x_2,\tau)$, $c_i=c(x_1,x_2,\tau)$,
$d=d(x_1,x_2,\tau)$ map $Q_T$ to
$\mathbb{R}$, and $a_i\left( \cdot , \tau\right)$, $b\left( \cdot , \tau\right)$, $c_i\left( \cdot , \tau\right)$, and $d\left( \cdot , \tau\right)$ are assumed to be in $C^2(\Omega)$ 
and $u\left(\cdot , t\right) \in C^6(\Omega)$ for all $\tau\in ]0,T]$. 
We define a uniform spatial grid $G$ with step size $\Delta x_k$ in $x_k$ direction for $k=1,2$.
Setting $f=-du_{\tau}$ and applying a standard, second-order central
difference approximation leads to the elliptic problem
\begin{align}\label{semidiscretepdezudiskretisierenII_before_applying_uxxx_and_uyyy}
f =&
A_0 
- \frac{a_1(\Delta x_1)^2}{12} \frac{\partial^4 u}{\partial x_1^4} 
- \frac{a_2(\Delta x_2)^2}{12} \frac{\partial^4 u}{\partial x_2^4} 
- \frac{b_{12}(\Delta x_1)^2}{6} \frac{\partial^4 u}{\partial
     x_1^3\partial x_2} \nonumber \\
&- \frac{b_{12}(\Delta x_2)^2}{6} \frac{\partial^4 u}{\partial x_1\partial x_2^3} 
- \frac{c_1(\Delta x_1)^2}{6} \frac{\partial^3 u}{\partial x_1^3}
- \frac{c_2(\Delta x_2)^2}{6}\frac{\partial^3 u}{\partial x_2^3} 
+ \varepsilon,
\end{align}
with $A_0 := a_1 D^c_{1}D^c_{1} U_{i_1,i_2}  + a_2 D^c_{2}D^c_{2} U_{i_1,i_2} +
b_{12} D^c_{1}D^c_{2} U_{i_1,i_2}  + c_1D^c_{1}U_{i_1,i_2}  + c_2
D^c_{2} U_{i_1,i_2}$, where $D_k^c$  denotes the central difference operator in $x_k$ direction, and $\varepsilon \in \mathcal{O}\left( h^4\right)$ if $\Delta x_k\in \mathcal{O}\left(h\right)$ for $h>0$.
We call a finite difference scheme high-order compact (HOC) if
its consistency error is of order $\mathcal{O}\left(h^4\right)$ for $\Delta
x_1, \Delta x_2\in\mathcal{O}\left(h\right)$ for $h>0$, and it uses
only points on the compact stencil, $U_{k,p}$ with $k\in
\{i_1-1,i_1,i_1+1\}$ and $p\in\{i_2-1,i_2,i_2+1\}$, to approximate the
solution at $\left(x_{i_1},x_{i_2}\right)\in \interior{G}$.

\section{Auxiliary relations for higher derivatives}\label{sec:Aux_relations}

Our aim is to replace the third- and fourth-order derivatives in
\eqref{semidiscretepdezudiskretisierenII_before_applying_uxxx_and_uyyy}
which are multiplied by second-order terms by equivalent expressions
which can be approximated with second order on the compact
stencil. Indeed, if we differentiate
\eqref{Definition_of_general_grid_2D_EHOC_before_using_f} (using $f=
-du_\tau$) once with respect to $x_k$ ($k = 1,2$), we obtain relations
\begin{align}\label{aux_eq_third_derivatives}
\frac{\partial^3 u}{\partial x_1^3} =& A_{1}, \qquad \frac{\partial^3 u}{\partial x_2^3} = A_{2}, 
\end{align}
where we can discretise $A_i$ with second order on the compact stencil using the central difference operator. Analogously, we obtain
\begin{align}\label{aux_eq_fourth_derivatives}
\frac{\partial^4 u}{\partial x_1^4} =& B_{1} 
- \frac{b_{12}}{a_1} \frac{\partial^4 u }{\partial x_1^3 \partial x_2} 
 &\iff \qquad 
\frac{\partial^4 u }{\partial x_1^3 \partial x_2} &=  \frac{a_1}{b_{12}} B_1 - \frac{a_1}{b_{12}} \frac{\partial^4 u }{\partial x_1^4 },\nonumber
\\
\frac{\partial^4 u}{\partial x_2^4} =& B_{2} - \frac{ b_{12} }{a_2} \frac{\partial^4 u }{\partial x_1 \partial x_2^3} 
 &\iff \qquad  
\frac{\partial^4 u }{\partial x_1 \partial x_2^3} &= \frac{ a_2}{b_{12} } B_{2} - \frac{ a_2}{b_{12} } \frac{\partial^4 u }{\partial x_2^4} ,
\\
\frac{\partial^4 u}{\partial x_1^3 \partial x_2}  = & C_{1} - \frac{a_2}{a_1} \frac{\partial^4 u }{\partial x_1 \partial x_2^3}  
 &\iff \qquad  
 \frac{\partial^4 u }{\partial x_1 \partial x_2^3}&= 
C_{2} - \frac{a_1}{a_2} \frac{\partial^4 u }{\partial x_1^3 \partial x_2} ,\nonumber
\end{align}
where we can approximate $B_k$ and $C_k$ with second order on the compact stencil using the central difference operator. 
A detailed derivation can be found in \cite{DuHe15,He14}.

\section{Derivation of high-order compact schemes}\label{sec:derivation}
In general it is not possible to obtain a HOC scheme for \eqref{Definition_of_general_grid_2D_EHOC_before_using_f}, 
since there are four fourth-order derivatives in \eqref{semidiscretepdezudiskretisierenII_before_applying_uxxx_and_uyyy}, 
but only three auxiliary equations for these in
\eqref{aux_eq_fourth_derivatives}. Hence, we propose four different
versions of the numerical schemes, 
where only one of the fourth-order derivatives in \eqref{semidiscretepdezudiskretisierenII_before_applying_uxxx_and_uyyy} is left as a second-order remainder term.
Using \eqref{aux_eq_third_derivatives} and \eqref{aux_eq_fourth_derivatives} in \eqref{semidiscretepdezudiskretisierenII_before_applying_uxxx_and_uyyy} we obtain
as 
\textit{Version 1\/} scheme
\begin{align}\label{Version1}
\begin{split}
f =&
A_0 
- \frac{c_1(\Delta x_1)^2}{6} A_1 
- \frac{c_2(\Delta x_2)^2}{6}A_2  
- \frac{a_2(\Delta x_2)^2}{12}B_{2} 
- \frac{b_{12}(\Delta x_2)^2}{12}C_{2}
\\
& 
- \frac{a_1\left(2a_2(\Delta x_1)^2 - a_1 (\Delta x_2)^2\right)}{12a_2} B_1 
+ \frac{a_1\left(a_2(\Delta x_1)^2 - a_1 (\Delta x_2)^2\right)}{12a_2} \frac{\partial^4 u }{\partial x_1^4 }  
+ \varepsilon ,
\end{split}
\end{align}
as \textit{Version 2\/} scheme 
\begin{align}\label{Version2}
\begin{split}
f =&
A_0 
- \frac{c_1(\Delta x_1)^2}{6} A_1 
- \frac{c_2(\Delta x_2)^2}{6}A_2  
- \frac{a_1(\Delta x_1)^2}{12} B_{1} 
- \frac{b_{12}(\Delta x_1)^2}{12} C_{1} 
\\
& 
- \frac{a_2\left(2a_1(\Delta x_2)^2 - a_2(\Delta x_1)^2 \right)}{12a_1}  B_{2} 
+ \frac{a_2\left(a_1(\Delta x_2)^2 - a_2(\Delta x_1)^2 \right)}{12a_1} \frac{\partial^4 u }{\partial x_2^4} 
+ \varepsilon ,
\end{split}
\end{align}
as \textit{Version 3\/} scheme 
\begin{align}\label{Version3}
\begin{split}
f =&
A_0 
- \frac{c_1(\Delta x_1)^2}{6} A_1 
- \frac{c_2(\Delta x_2)^2}{6}A_2  
- \frac{a_1(\Delta x_1)^2}{12} B_{1} 
- \frac{a_2(\Delta x_2)^2}{12} B_{2}
\\
& 
- \frac{b_{12}(\Delta x_2)^2}{12} C_{2}
+ \frac{b_{12}\left(a_1(\Delta x_2)^2 - a_2(\Delta x_1)^2 \right)}{12a_2} \frac{\partial^4 u }{\partial x_1^3 \partial x_2}
+ \varepsilon ,
\end{split}
\end{align}
and, finally, as \textit{Version 4\/} scheme
\begin{align}\label{Version4}
\begin{split}
f =&
A_0 
- \frac{c_1(\Delta x_1)^2}{6} A_1 
- \frac{c_2(\Delta x_2)^2}{6}A_2  
- \frac{a_1(\Delta x_1)^2}{12} B_{1} 
- \frac{a_2(\Delta x_2)^2}{12} B_{2}
\\
& 
- \frac{b_{12}(\Delta x_1)^2}{12} C_{1} 
+ \frac{b_{12}\left(a_2(\Delta x_1)^2 - a_1(\Delta x_2)^2 \right)}{12a_1} \frac{\partial^4 u }{\partial x_1 \partial x_2^3} 
+ \varepsilon .
\end{split}
\end{align}
Employing the central difference operator with $\Delta x=\Delta y =h$
for $h>0$ to discretise $A_i,$ $B_i,$ 
$C_i,$ in \eqref{Version1}--\eqref{Version4} and neglecting the
remaining lower-order term leads to four semi-discrete (in space)
schemes. A more detailed 
description of this approach can be found in \cite{DuHe15,He14}. 
When $a_1 \equiv a_2$ or $b_{12}\equiv 0$ these schemes are
fourth-order consistent in space, otherwise second-order. 

In time, we apply the implicit BDF4 method on an equidistant time grid with stepsize $k \in \mathcal{O}\bigl(h\bigr)$.
The necessary starting values are obtained using a Crank-Nicolson time discretisation, 
where we subdivide the first timesteps with a step size $k' \in \mathcal{O}\left(h^2\right)$ to ensure the fourth-order time discretisation in terms of $h$. 

With additional information on the solution of \eqref{Definition_of_general_grid_2D_EHOC_before_using_f} even better results are possible. 
If the specific combination of pre-factors in \eqref{Definition_of_general_grid_2D_EHOC_before_using_f} and 
the higher derivatives in the second-order terms is sufficiently small, the second-order term dominates the computational error only for very small step-sizes $h$. 
Before this error term becomes dominant one can observe a fourth-order
numerical convergence.  In this case we call the scheme essentially
high-order compact (EHOC). 

\section{Application to option pricing}\label{Application_of_EHOC_schemes_to_Heston_with_zoom}
In this section we apply our numerical schemes to an option pricing
PDE in a family of
stochastic volatility models, with a generalised square root process
for the variance with nonlinear drift term,
\begin{align*}
dS_t = &\mu S_t {\rm d}t + \sqrt{v_t}S_t{\rm d}W^{(1)}_t,\qquad
dv_t = \kappa v_t^\alpha \left(\theta -v_t \right)  {\rm d}t + \sigma \sqrt{v_t} {\rm d}W^{(2)}_t ,
\end{align*}
with $\alpha\geq 0$, a correlated, two-dimensional Brownian motion, $dW_t^{(1)}dW_t^{(2)}=\rho
dt$, as well as drift $\mu\in \mathbb{R}$ of the stock price $S$, long
run mean $\theta>0$, mean reversion speed $\kappa>0$, and volatility of volatility $\sigma>0$.
For $\alpha=0$ one obtains the standard Heston model, for $\alpha = 1$ the SQRN model, see \cite{ChJaMi10}.
Using It{\^o}'s lemma and standard arbitrage arguments, the option
price $V=V(S,v,t)$ solves
\begin{equation}\label{pde_general_SQR_model_untransformed}
\frac{\partial V}{\partial t} 
+ \frac{vS^2}{2}\frac{\partial^2 V}{\partial S^2} 
+ \rho \sigma v S\frac{\partial^2 V}{\partial S \partial v} 
+ \frac{\sigma^2 v}{2}\frac{\partial^2 V}{\partial v^2}
 +rS\frac{\partial V}{\partial S} 
 + \kappa v^\alpha \left(\theta -v\right) \frac{\partial V}{\partial v} 
 -rV=0 ,
\end{equation}
where $S,\sigma>0$ and $t \in \left[0,T\right[$ with $T>0$.
For a European Put with exercise price $K$ we have the final condition
$ V(S,T) = \max\left( K-S,0\right)$. The transformations 
$\tau = T-t$, $u = e^{r\tau}{V}/{K},$ $\hat{S}=\ln (  {S}/{K} ),$ $y= {v}/{\sigma} $
as well as $\hat{S}=\varphi\left(x\right)$ \cite{DuFoHe14}, lead to
\begin{multline*}\label{pdezudiskretisierenII}
 \varphi_x^3 u_{\tau} 
 +\frac{\sigma y}{2}\left[\varphi_x u_{xx}  + \varphi_x^3 u_{yy}\right] 
 - \rho \sigma y\varphi_x^2 u_{xy} \\
 + \left[\frac{\sigma y\varphi_{xx}}{2} + \Bigl(\frac{\sigma y}{2}-r\Bigr)\varphi_x^2 \right] u_x 
 - \kappa \sigma^\alpha y^\alpha\frac{\theta -\sigma y}{\sigma}\varphi_x^3 u_y =0,
\end{multline*}
with initial condition $u(x,y,0)= \max \left(1-e^{\varphi(x)},0\right)$. 
The function $\varphi$ is considered to be four times differentiable
and strictly monotone. It is chosen in
such a way that grid points are concentrated around the exercise price
$K$ in the $S$\nobreakdash--$v$~plane when using a uniform grid in the $x$\nobreakdash--$y$~plane.

Dirichlet boundary conditions are imposed at $x=x_{\min} $ and $x=x_{\max}$ similarly as in \cite{DuFoHe14},
\begin{equation*}
u(x_{\min},y,\tau)=u(x_{\min},y,0),  \quad u(x_{\max},y, \tau)= u(x_{\max},y, 0),
\end{equation*}
for all $\tau \in [0,\tau_{\max}]$ and $y \in [y_{\min},y_{\max}].$ 
At the boundaries
${y=y_{\min}}$ and ${y=y_{\max}}$ we employ the discretisation of the
interior spatial domain and extrapolate the resulting ghost-points  
using 
\begin{align*}
U_{i,-1}&= 3 U_{i,0} - 3 U_{i,1} + U_{i,2} + \mathcal{O}\bigl(h^3
          \bigr), \\
 U_{i,M+1} &= 3 U_{i,M} - 3 U_{i,M-1} + U_{i,M-2} + \mathcal{O}\bigl (h^3 \bigr),
\end{align*}
for $i=0, \ldots , N$. Third-order extrapolation is sufficient here to
ensure overall fourth-order convergence \cite{Gus81}.

\section{Numerical experiments}\label{Application_of_Version_3_Heston_with_zoom} 
We employ the function
$\varphi(x)= \sinh(c_2x+c_1(1-x))/\zeta,$
where $c_1= \asinh(\zeta \hat{S}_{\min})$,
$ c_2= \asinh(\zeta \hat{S}_{\max})$ and $\zeta>0$.
We use  $\kappa = 1.1$, $\theta = 0.2$, $v = 0.3$, $r=0.05$, $K=100$, $T=0.25$, $v_{\min} =0.1$, $v_{\max}=0.3$, $S_{\min} = 1.5$, $S_{\max} = 250$, $\rho = 0, -0.4$ and $\zeta = 7.5$. 
Hence, $x_{\max} - x_{\min}=y_{\max} - y_{\min}=1 $.
For the Crank-Nicolson method we use $k'/h^2=0.4$, for the BDF4 method $k/h = 0.1$. We smooth the initial condition according to \cite{KrThWi70,DuHe15}, 
so that the smoothed initial condition tends towards the original initial condition for $h\rightarrow 0$. 
We neglect the case $\alpha=0$ (Heston model), since a numerical study of that case has been performed in \cite{DuFoHe14}.
In the numerical convergence plots we use a reference solution
$U_{\text{ref}}$ on a fine grid ($h = 1/320$) and report the absolute $l^2$-error compared to $U_{\text{ref}}$.
The numerical convergence order is computed from the slope of the
linear least square fit of the points in the log-log plot. 
\begin{figure}[h!]\centering
\subfigure[Transformation with $\zeta=7.5$]{\includegraphics[width=0.3\linewidth]{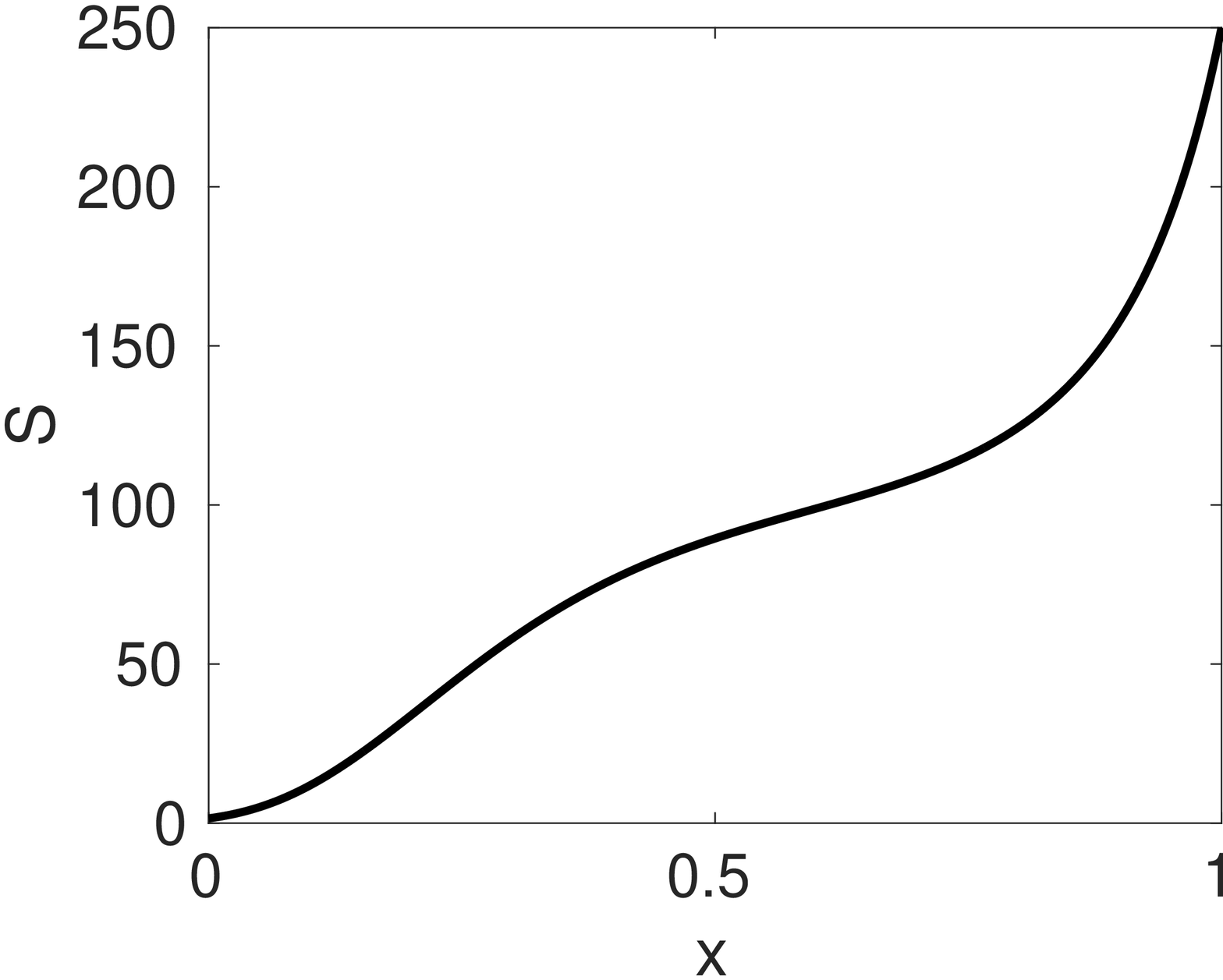}\label{fig:zoom}}
\subfigure[HOC: $\alpha=0.25$, $\rho=0$]{\includegraphics[width=0.3\linewidth]{SV_model_family_absolute_L2_error_K=100_a=0_25_b=0_5_rho=0_zeta=7_5.eps}\label{fig:numconv1}}
\subfigure[HOC: $\alpha=0.5$, $\rho=0$]{\includegraphics[width=0.3\linewidth]{SV_model_family_absolute_L2_error_K=100_a=0_5_b=0_5_rho=0_zeta=7_5.eps}\label{fig:numconv2}}\\
\subfigure[HOC: $\alpha=0.75$, $\rho=0$]{\includegraphics[width=0.3\linewidth]{SV_model_family_absolute_L2_error_K=100_a=0_75_b=0_5_rho=0_zeta=7_5.eps}\label{fig:numconv3}}
\subfigure[HOC: SQRN model, $\rho=0$]{\includegraphics[width=0.3\linewidth]{SV_model_family_absolute_L2_error_K=100_a=1_b=0_5_rho=0_zeta=7_5.eps}\label{fig:numconv4}}
\subfigure[EHOC: $\alpha=0.25$, $\rho=-0.4$]{\includegraphics[width=0.3\linewidth]{SV_model_family_absolute_L2_error_K=100_a=0_25_b=0_5_rho=-0_4_zeta=7_5.eps}}\\
\subfigure[EHOC: $\alpha=0.5$, $\rho=-0.4$]{\includegraphics[width=0.3\linewidth]{SV_model_family_absolute_L2_error_K=100_a=0_5_b=0_5_rho=-0_4_zeta=7_5.eps}}
\subfigure[EHOC: $\alpha=0.75$, $\rho=-0.4$]{\includegraphics[width=0.3\linewidth]{SV_model_family_absolute_L2_error_K=100_a=0_75_b=0_5_rho=-0_4_zeta=7_5.eps}}
\subfigure[EHOC: SQRN model, $\rho=-0.4$]{\includegraphics[width=0.3\linewidth]{SV_model_family_absolute_L2_error_K=100_a=1_b=0_5_rho=-0_4_zeta=7_5.eps}}
\caption{Transformation of the spatial grid and numerical convergence plots.}
\end{figure}

Figure~\ref{fig:zoom} shows the transformation from $x$ to $S$. The transformation focuses on the region around the strike price. 
Figures~\ref{fig:numconv1}, \ref{fig:numconv2}, \ref{fig:numconv3} and \ref{fig:numconv4} show that the HOC
schemes lead to a numerical convergence order of about $3.5$, whereas
the standard, second-order central difference discretisation (SD)
leads to convergence orders of about $2.3$, in the case of vanishing correlation.
In all cases with non-vanishing correlation ($\rho\neq 0$) we observe only slightly improved convergence for Version 1 (V1) when comparing it to the standard discretisation.
Version 2 (V2) and Version 3 (V3), however, lead to similar convergence orders as the HOC scheme, even for non-vanishing correlation. 
Results of Version 4 are not shown as this scheme shows instable
behaviour in this example.

In summary, we obtain high-order compact schemes for vanishing
correlation and achieve high-order convergence also for non-vanishing correlation for the family \eqref{pde_general_SQR_model_untransformed} of stochastic volatility model. A standard, second-order discretisation is significantly outperformed in all cases.

\subsection*{Acknowledgment}
\noindent 
BD acknowledges support by the Leverhulme Trust research project grant
`Novel discretisations for higher-order nonlinear PDE' (RPG-2015-69).
CH was supported by the European Union in the FP7-PEOPLE-2012-ITN Program under Grant Agreement Number 304617 (FP7 Marie Curie Action, 
Project Multi-ITN STRIKE -- Novel Methods in Computational Finance).


\begin{thebibliography}{9}
\bibitem{ChJaMi10}
P.~Christoffersen, K.~Jacobs and K.~Mimouni.
\newblock Models for {S}\&{P}500 dynamics: evidence from realized volatility, daily returns, and option prices.
\newblock {\em Rev.~Financ.~Stud.}, 23(8):3141--3189, 2010.

\bibitem{DuFoHe14}
B.~D{\"u}ring, M.~Fourni{\'e} and C.~Heuer.
\newblock High-order compact finite difference schemes for option pricing in
  stochastic volatility models on non-uniform grids.
\newblock {\em J.~Comput.~Appl.~Math.}, 271(18):247--266, 2014.

\bibitem{DuHe15} B.~D{\"u}ring and C.~Heuer.
\newblock High-order compact schemes for parabolic problems with mixed
  derivatives in multiple space dimensions.
\newblock {\em SIAM J. Numer. Anal.}, 53(5):2113--2134, 2015.

\bibitem{Gus81}
B.~Gustafsson.
\newblock The convergence rate for difference approximations to general mixed
  initial-boundary value problems.
\newblock {\em SIAM~J.~Numer.~Anal.}, 18(2):179--190, 1981.

\bibitem{He14}
C.~Heuer.
\newblock {\em High-order compact finite difference schemes for parabolic
  partial differential equations with mixed derivative terms and applications
  in computational finance}.
\newblock PhD thesis, University of Sussex, August 2014. \url{http://sro.sussex.ac.uk/49800/}

\bibitem{KrThWi70}
H.O. {Kreiss}, V.~{Thomee} and O.~{Widlund}.
\newblock {Smoothing of initial data and rates of convergence for parabolic
  difference equations.}
\newblock {\em {Commun.~Pure~Appl.~Math.}}, 23:241--259, 1970.

\end{thebibliography}
\end{document}